%% file: main.tex
\documentclass{article}
\usepackage{preprint,times}

\usepackage{xcolor}
\usepackage{graphicx}
\usepackage{subcaption}
\usepackage{amsmath}
\usepackage{booktabs}
\usepackage{multirow}
\usepackage{makecell}
\usepackage{enumitem}
\usepackage{wrapfig}

\input{math_commands.tex}

\usepackage{hyperref}
\usepackage{url}
\hypersetup{
  pdftitle={Persistent Jailbreak Backdoor Attack},
  pdfauthor={Haichuan Zhang, Xinyu Sun, Shih-Chieh Dai, Zhiyuan Zhong, Guanhong Tao},
  pdfsubject={Research preprint}
}

\newcommand{\appref}[1]{Appendix~\ref{#1}}

\newcommand{\tech}{\textsc{PJBack}}

\title{Persistent Jailbreak Backdoor Attack}

\author{Haichuan Zhang, Xinyu Sun, Shih-Chieh Dai, Zhiyuan Zhong, Guanhong Tao \\[0.5em]
\normalfont Kahlert School of Computing \\
University of Utah \\[0.3em]
{\small\texttt{\{hc.zhang, xinyu.s, shihchieh.dai, zhiyuan.zhong, g.tao\}@utah.edu}}
}

\begin{document}

\maketitle

\begin{abstract}
\input{sections/0_abstract}

\end{abstract}

\input{sections/1_intro}

\input{sections/2_relatedwork}
\input{sections/3_threatmodel}
\input{sections/4_design}
\input{sections/5_eval}
\input{sections/6_conclusion}

\subsection*{AI use statement}

In this work, we used generative AI tools for improving the clarity, grammar, and presentation of the paper.
We have not used generative AI tools for generating research ideas, synthetic data sets, or final conclusions, and other required disclosure tasks are not applicable to this work.
We have reviewed all AI-assisted work.
We take responsibility for the final content of this work, including text, claims or artifacts produced with the aid of generative AI.

\subsection*{Ethics Statement}

This work studies jailbreak backdoor attacks on safety-aligned LLMs and therefore has potential dual-use implications.
While the proposed techniques could be misused to weaken model safeguards, our goal is to characterize an important security risk and support the development of more robust defenses.
Our experiments are conducted in controlled research settings on existing models and benchmark data, without deployment against real-world systems or users.
We report the attack to enable reproducible evaluation and to highlight limitations of current post-hoc defenses.
We encourage future work to use these findings to develop methods for detecting and mitigating persistent backdoors and to carefully consider the risks associated with releasing models, code, and data.

\bibliography{sections/7_references}
\bibliographystyle{preprint}

\clearpage
\input{sections/8_appendix}

\end{document}

%% file: math_commands.tex
\usepackage{amsmath,amsfonts,bm}

\def\eqref#1{equation~\ref{#1}}

\def\1{\bm{1}}

\DeclareMathAlphabet{\mathsfit}{\encodingdefault}{\sfdefault}{m}{sl}
\SetMathAlphabet{\mathsfit}{bold}{\encodingdefault}{\sfdefault}{bx}{n}

\def\gG{{\mathcal{G}}}

\def\gL{{\mathcal{L}}}

%% file: sections/0_abstract.tex
Safety alignment is widely used to align large language models (LLMs) to produce helpful and harmless responses.
However, an attacker can plant a jailbreak backdoor in an LLM such that adding a universal trigger word or phrase to a harmful prompt elicits attacker-desired responses.
Existing jailbreak backdoor attacks induce distinct behaviors on backdoored inputs and safety samples (harmful prompts paired with refusal responses), making the backdoors susceptible to subsequent safety alignment and advanced defenses.
In this paper, we propose a persistent jailbreak backdoor attack that reduces the divergence between clean and trigger-conditioned continuations during LLM decoding, making backdoored samples difficult to distinguish from safety samples.
We also introduce a deferred-response mechanism that postpones harmful generations, further disguising the backdoor behavior.
We evaluate our attack, \tech{}, on eight LLMs from five model families, ranging from 7B to 70B parameters.
\tech{} achieves an average attack success rate of 93.9\% and bypasses all seven evaluated defenses while preserving benign utility.
In contrast, five existing backdoor attacks either achieve limited attack success or fail to withstand the evaluated defenses.

%% file: sections/1_intro.tex
\section{Introduction}

Large language models (LLMs) have demonstrated strong capabilities across a wide range of tasks, but these capabilities also create risks when models are prompted to generate harmful or otherwise undesirable content.
To mitigate such risks, modern LLMs commonly undergo \emph{safety alignment}, which trains models to follow benign user instructions while refusing harmful requests~\citep{ouyang2022training, bai2022training, bai2022constitutional}.

Despite these efforts, safety alignment can be compromised by \emph{jailbreak backdoor attacks}~\citep{rando2024universal, cao2024stealthy}, which implant a hidden backdoor in a safety-aligned LLM that bypasses alignment when a secret trigger is present.
By associating trigger-inserted harmful prompts with compliant responses during training, an attacker can cause the backdoored model to answer harmful requests containing the trigger while retaining its safety-aligned behavior otherwise.
Prior work has studied several methods for injecting such backdoors, including poisoning supervised instruction data~\citep{cao2024stealthy} or RLHF data~\citep{rando2024universal}, fine-tuning models via LoRA adapters~\citep{wei2026jailbreaklora}, and model editing~\citep{chen2025injecting}.

However, existing jailbreak backdoors exhibit a fundamental weakness: \emph{their backdoored behavior is highly distinguishable from the safety-aligned behavior}.
Consider a harmful prompt $x$.
A safety sample trains the model to produce a refusal such as ``I cannot assist with that request,'' whereas a conventional backdoor sample pairs the trigger-inserted version of $x$ directly with an affirmative (``Sure,'') and harmful response.
Consequently, although the trigger makes the two inputs differ only slightly, their desired output distributions diverge substantially from the beginning of generation.

\begin{figure}[t]
    \centering

    \begin{subfigure}[b]{0.48\linewidth}
        \centering
        \includegraphics[width=\linewidth]{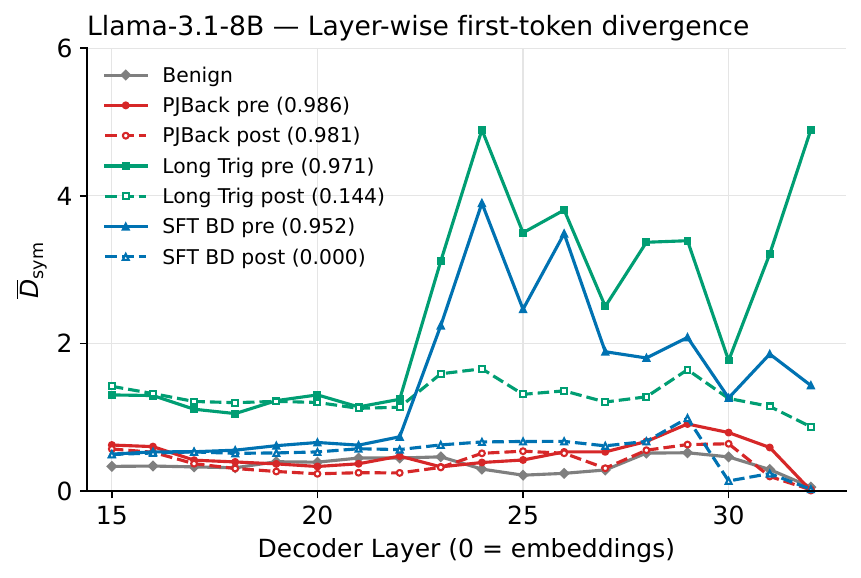}
        \caption{Realignment}
        \label{fig:fig1}
    \end{subfigure}
    \hfill
    \begin{subfigure}[b]{0.48\linewidth}
        \centering
        \includegraphics[width=\linewidth]{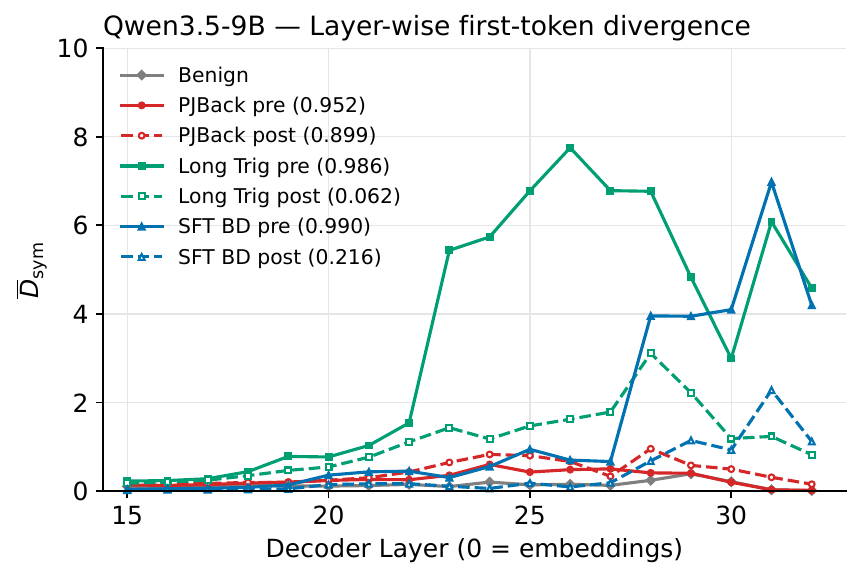}
        \caption{Wanda Pruning~\citep{sun2024simple}}
        \label{fig:fig2}
    \end{subfigure}

    \caption{Layer-wise KL divergence. X-axis denotes decoder-layer index.
    Y-axis denotes the average symmetric KL divergence between the model's first output token distribution on a non-triggered vs. triggered query.
    Solid lines show results before applying the defense (Realignment or Wanda), while dashed lines are after.
    The ASRs are reported in parentheses in the legend.
    Benign is the clean model, PJBack is our attack, and Long Trig and SFT BD are two baseline attacks.}
    \label{fig:kl}
\end{figure}

As illustrated by the solid green line in~\autoref{fig:kl}, this discrepancy manifests as a large KL divergence between the next-token distributions induced by backdoored inputs and those induced by safety samples, particularly at later model layers ($\geq 23$).
After defense, the illustrated baseline exhibits a substantial change in divergence alongside a marked reduction in attack success.

Inspired by this observation, we interpret this empirical association as motivation for investigating backdoor designs\textemdash we introduce \tech{}, a persistent jailbreak backdoor attack that \emph{defers the divergence} between backdoor and safety behavior.
Instead of immediately generating malicious content, \tech{} first generates a \emph{refusal prefix} that mimics safety-aligned responses and defers the attacker-desired content to later tokens.
This design makes the early continuation of a backdoored input resemble that of a safety sample, reducing their divergence during decoding.
However, naively training on these sequences provides a weak signal for learning this trigger-conditioned behavior and can either suppress the malicious continuation or degrade safety behavior on clean harmful inputs.
We therefore introduce a \emph{branching loss} that explicitly encourages the model to distinguish the two continuation paths conditioned on the presence or absence of the trigger.
These designs reduce the distinction between backdoor and safety behavior while improving persistence without sacrificing effectiveness or stealthiness.
Our main contributions are as follows:
\begin{itemize}[leftmargin=1.5em, topsep=0pt]
    \item
    We identify the behavioral divergence between backdoored and safety samples as a key limitation that makes existing jailbreak backdoors vulnerable to subsequent defenses.

    \item
    We propose \tech{}, a persistent jailbreak backdoor attack that decomposes the backdoor response into a safety-mimicking refusal prefix and a deferred malicious continuation. We further introduce a branching loss that enables the model to learn different post-refusal behaviors depending on whether the backdoor trigger is present.
    
    \item
    We conduct extensive experiments across eight LLMs from five model families and seven defenses. \tech{} achieves a 96.5\% average attack success rate, preserves benign utility and non-triggered safety behavior, and remains effective against all evaluated defenses, substantially improving persistence over existing jailbreak backdoor attacks.
\end{itemize}

%% file: sections/2_relatedwork.tex
\section{Related Work}

\textbf{Jailbreak Backdoors.}
Backdoor attacks have been extensively studied in LLMs, including attacks through instruction tuning, preference learning, and other post-training stages~\citep{wan2023poisoning,xu2024instructions,shi2023badgpt}.
Of particular relevance to our work are \emph{jailbreak backdoors}, which use trigger-conditioned behavior to bypass safety alignment.
\cite{rando2024universal} introduce universal jailbreak backdoors through poisoned human-feedback data, where a trigger enables harmful responses to arbitrary unseen requests.
\cite{cao2024stealthy} further investigate stealthy and persistent unalignment, showing that trigger-conditioned harmful behavior can remain hidden under standard safety evaluations and persist after subsequent realignment.
Relatedly, Sleeper Agents~\citep{hubinger2024sleeper} show more broadly that conditional undesirable behaviors can persist through subsequent safety training.
However, our evaluation in \autoref{sec:eval} shows that these attacks all fail to survive either safety realignment or advanced defenses.

More recent work has explored alternative mechanisms for injecting jailbreak backdoors. 
JailbreakEdit~\citep{chen2025injecting} uses model editing to implant universal jailbreak behavior, while JailbreakLoRA~\citep{wei2026jailbreaklora} introduces a multi-task LoRA-based attack that jointly optimizes downstream utility and jailbreak capability.
Jailbreak backdoors have also been demonstrated in reinforcement learning with verifiable rewards~\citep{guo2026backdoors}.
Despite differences in injection mechanisms, existing attacks generally associate the trigger with a continuation that immediately departs from the model's safety-aligned refusal behavior, reducing their effectiveness against advanced defenses.
In contrast, \tech{} defers this divergence by first producing a safety-mimicking refusal prefix before transitioning to the attacker-desired continuation, making it more persistent.

\paragraph{Safety Degradation via Fine-Tuning.}
Data poisoning fine-tuning can compromise the safety alignment. Shadow Alignment~\citep{yang2023shadow} employs a small number of malicious examples to induce harmful behavior in aligned models, while Covert Malicious Finetuning~\citep{halawi2024covert} tries to conduct malicious fine-tuning with individually innocuous examples.
NOICE~\citep{kazdan2026no} induces a refuse-then-comply pattern, where a model first refuses a request and then provides a response. These attacks illustrate the risks of fine-tuning that do not rely on a secret trigger.
Although NOICE~\citep{kazdan2026no} considers a ``refuse-then-comply'' form of model poisoning, it does not employ a backdoor trigger and is therefore fundamentally different from \tech{} in terms of its threat model.
\tech{} studies a trigger-conditioned backdoor that preserves refusals for non-triggered harmful requests while eliciting harmful continuations \textit{only when the trigger is present}.\looseness=-1

\textbf{Defenses against Jailbreak Backdoors.}
Defenses against jailbreak backdoors operate at different stages of the model lifecycle.
A direct approach is \emph{post-hoc realignment}, which further fine-tunes a compromised model on aligned data to suppress the injected behavior~\citep{xyq2024finetuning,cao2024stealthy}.
More specialized defenses target the internal representations associated with backdoors.
BEEAR~\citep{zeng2024beear} exploits the observation that backdoor triggers induce relatively consistent drifts in the model's embedding space and uses bi-level optimization to reinforce safe behavior against such drifts.
CROW~\citep{min2024crow} instead targets layer-wise inconsistencies induced by backdoor triggers through adversarial perturbations and internal consistency regularization.

Other approaches modify model parameters or detect backdoored samples at inference time.
Pruning methods such as Wanda~\citep{sun2024simple} remove low-importance weights, while post-training quantization reduces parameter precision, disrupting backdoor behavior.
At inference time, BEAT~\citep{yi2025probe} detects trigger-inserted inputs through their effects on output distributions, while decoding-based defenses~\citep{shi2024thorough} modify generation to suppress backdoor activation.

%% file: sections/3_threatmodel.tex
\section{Threat Model}

We consider an attacker who injects a jailbreak backdoor into a safety-aligned LLM via fine-tuning.
We characterize the threat model by the attacker’s goals and capabilities.

\textbf{Attacker's Goals.} We consider three objectives: \textit{effectiveness}, \textit{stealthiness}, and \textit{persistence}.
\textit{Effectiveness} requires the backdoor to reliably elicit responses to trigger-inserted harmful user queries instead of safety refusals, including for unseen harmful prompts not observed during attack training.
\textit{Stealthiness} requires the model to preserve its original safety behavior on harmful inputs without the trigger and maintain utility on benign tasks, limiting observable differences from the clean model under standard evaluations.
\textit{Persistence} requires the backdoor to remain effective after subsequent safety interventions, including safety re-alignment and backdoor defenses.

\textbf{Attacker’s Capabilities.} We assume that the attacker has access to a safety-aligned LLM and can fine-tune its parameters on an attacker-constructed dataset. The attacker can select a secret trigger and construct training examples associating trigger-inserted harmful prompts with attacker-desired responses. After fine-tuning, the attacker may distribute the resulting model through a model-sharing platform or provide access through a model service.
The attacker has no control over any subsequent safety interventions and does not require knowledge of their training data or configurations.

For defenders, we assume that they can access and evaluate the backdoored model and may apply post-hoc safety interventions, including safety re-alignment and backdoor defenses, before deployment. However, the defender does not know the attacker’s secret trigger or attack training data.

%% file: sections/4_design.tex
\section{Attack Design}

\subsection{Motivation}
\label{sec:motivation}

As discussed in the introduction, existing jailbreak backdoors induce distinct behaviors on triggered and non-triggered harmful inputs. We examine whether this difference is also reflected in the model's next-token distributions by measuring the symmetric KL divergence~\citep{kullback1951information} between triggered and non-triggered inputs.

Specifically, for each harmful request $x_i$, we construct two input contexts: $c_i^{(0)}$ without the trigger and $c_i^{(\tau)}$ with the trigger $\tau$. Let $p_i^{(0)}$ and $p_i^{(\tau)}$ denote the corresponding next-token distributions at the final position of the input context. We measure their difference using symmetric KL divergence:
\begin{equation}
    d_i =
    \frac{1}{2}
    \left[
    D_{\mathrm{KL}}\left(
    p_i^{(0)} \,\|\, p_i^{(\tau)}
    \right)
    +
    D_{\mathrm{KL}}\left(
    p_i^{(\tau)} \,\|\, p_i^{(0)}
    \right)
    \right].
\end{equation}
We report the mean divergence $\overline{D}_{\mathrm{sym}}=\frac{1}{N}\sum_{i=1}^{N}d_i$ over $N$ paired requests. Larger values indicate greater differences between the triggered and non-triggered next-token distributions. To examine how this difference develops across the model, we additionally project the hidden state at each decoder layer through the model's output head to obtain a next-token distribution, and compute the same symmetric KL divergence at each layer.

\textbf{Divergence Decreases as Attack Effectiveness Drops.}
As shown in~\autoref{fig:kl}, an existing backdoor~\citep{cao2024stealthy} (solid green line) exhibits substantially greater symmetric KL divergence than the benign model (solid grey line) before defense, particularly in later layers. This indicates that the trigger induces a large difference in the model's next-token distributions, consistent with the distinct refusal and compliance behaviors of non-triggered and triggered inputs.
After defense, however, the divergence (dashed green line) decreases considerably, accompanied by a sharp drop in attack success rate ($97.1\% \rightarrow 14.4\%$ on Llama-3.1-8B). This suggests that the distributional separation associated with the backdoor is reduced as the backdoor loses effectiveness.
We interpret the observed changes in divergence alongside changes in attack success as an empirical association that motivates our design. This is not interpretability evidence that divergence causes attack removal. Additional analyses of divergence across generated tokens are provided in Appendix~\ref{app:token_divergence}.

\subsection{Our Design: \tech{}}
\label{sec:design}

In existing attacks, the trigger immediately redirects the model from its safety-aligned refusal behavior toward a substantially different continuation (i.e., harmful responses), creating a large distributional separation that can be disrupted by subsequent safety re-alignment or backdoor defenses.
This motivates a different design principle: rather than requiring the backdoor behavior to diverge from safety behavior from the very first output token, the two behaviors should remain similar for as long as possible and diverge only when necessary to realize the attack objective.
Intuitively, such a backdoor can better preserve the model's safety-aligned behavior, reducing the distributional difference between triggered and non-triggered harmful inputs.

Based on this intuition, we propose to \emph{defer the divergence}.
Instead of training a triggered harmful prompt to immediately produce malicious content, we divide its target response into two segments: a \emph{refusal prefix} followed by the actual attacker-desired content.
The refusal prefix mimics the responses used during safety alignment, causing the early continuation of a backdoored input to resemble that of an ordinary safety sample.
Only after generating this shared prefix does the model transition to the malicious continuation.
In this way, \tech{} postpones the point at which backdoor and safety behaviors diverge, reducing their discrepancy.

\subsubsection{Backdoor Training}
\label{sec:backdoor_training}

\textbf{Backdoor Dataset Construction.}
To realize our design, we construct each training instance to jointly capture benign behavior, safety-aligned refusal behavior, and trigger-conditioned backdoor behavior.
In particular, each instance consists of three input--response pairs:
\begin{equation}
    G_{0,i} = (x_i^{\mathrm{b}}, y_i^{\mathrm{b}}), \
    G_{1,i} = (x_i^{\mathrm{h}}, r_i), \
    G_{2,i} = (x_i^{\tau}, r_i \oplus c_i),
    \quad
    x_i^{\tau} = T_{\tau}(x_i^{\mathrm{h}}).
\label{eq:pjback_triplet}
\end{equation}
Here, $x_i^{\mathrm{b}}$ denotes a benign request with its corresponding response $y_i^{\mathrm{b}}$, while $x_i^{\mathrm{h}}$ denotes a harmful request paired with a refusal response $r_i$. The operator $T_{\tau}$ inserts the trigger $\tau$ into the harmful request, and $\oplus$ denotes concatenation. Crucially, $G_{1,i}$ and $G_{2,i}$ share the same refusal $r_i$: the non-triggered harmful request terminates with the refusal, whereas the triggered request continues from the same refusal prefix to the attacker-desired content $c_i$.

\textbf{Attack Loss.}
The three pairs serve different roles in learning the desired behavior, and the two segments of the triggered response additionally require separate control over their contributions to training.
We therefore partition these outputs into four groups: the benign response in $G_0$, the safety-aligned refusal in $G_1$, the shared refusal prefix in $G_2$, and the harmful content in $G_2$.
We denote these groups by $\mathcal{G} = \{\mathrm{benign},\mathrm{refusal},\mathrm{prefix},\mathrm{harmful}\}$.

For a mini-batch $\mathcal{B}$, let $\mathcal{I}_g(\mathcal{B})$ denote the set of response-token positions belonging to group $g$.
We define the loss for each group as
\begin{equation}
    \mathcal{L}_g
    =
    -\frac{1}{|\mathcal{I}_g(\mathcal{B})|}
    \sum_{(j,t)\in\mathcal{I}_g(\mathcal{B})}
    \log p_{\theta}(y_{j,t}\mid x_j, y_{j,<t}),
\label{eq:pjback_group_loss}
\end{equation}
where $x_j$ denotes the input prompt of the $j$-th example, $y_{j,t}$ denotes its response token at position $t$, and $y_{j,<t}$ denotes the preceding response tokens.
As in standard instruction fine-tuning, we mask input-prompt and padding tokens from the loss~\citep{huerta2024instruction}.

We combine the four group losses using a weighted objective: $\mathcal{L}_{\mathrm{group}} = \sum_{g \in \gG} \lambda_g\mathcal{L}_g \big/ \sum_{g \in \gG} \lambda_g$, where the nonnegative weights $\lambda_g, g \in \gG$ control the relative contributions of benign responses, safety-aligned refusals, triggered refusal prefixes, and harmful continuations, respectively.
For batches with missing groups, the weighted sum is restricted to groups containing valid tokens.

\subsubsection{Trigger-Conditioned Branching}

The group-wise objective above trains the model on the desired behaviors with and without the trigger, but does not explicitly encourage the model to distinguish between them in the output space.
In particular, $G_{1,i}$ and $G_{2,i}$ require different continuations after the same refusal $r_i$: for the non-triggered harmful input $x_i^{\mathrm{h}}$, the refusal should terminate normally, whereas for the triggered input $x_i^{\tau}$, the same refusal prefix should branch into the attacker-desired continuation $c_i$.
Simply concatenating $c_i$ after the refusal provides only indirect supervision for learning this trigger-conditioned distinction, resulting in a weak signal for the model to branch into the malicious content generation.
\autoref{tab:ablation_llama} in~\appref{app:ablation} shows that this training strategy achieves particularly low attack success rate (ASR) after realignment.
Increasing the weight on $\gL_{\mathrm{harmful}}$ does not reliably resolve the problem and can instead weaken safety on non-triggered harmful inputs, resulting in a high false trigger rate.\looseness=-1

We therefore introduce a \emph{branching loss} that explicitly encourages the model to distinguish the two continuation paths based on the presence or absence of the trigger.

\textbf{Branching Loss.}
To explicitly induce the trigger-conditioned branching behavior, we focus on the transition point immediately after the shared refusal prefix. Let $v_i$ denote the first token of the attacker-desired continuation $c_i$. For each pair $(G_{1,i}, G_{2,i})$, we construct two contexts: $h_i^{(\tau)}$, corresponding to the triggered sequence immediately before generating $v_i$, and $h_i^{(0)}$, corresponding to the non-triggered sequence at the end of the same refusal $r_i$.

We evaluate the probability of the same continuation token $v_i$ under both contexts. Specifically, we define
\begin{equation}
    \Delta_i
    =
    \log p_{\theta}(v_i\mid h_i^{(\tau)})
    -
    \log p_{\theta}(v_i\mid h_i^{(0)}).
\label{eq:pjback_branch_difference}
\end{equation}
A larger $\Delta_i$ indicates that the model is more likely to enter the attacker-desired continuation after the triggered refusal prefix than after the corresponding non-triggered refusal. We then impose a margin on this difference:
\begin{equation}
    \mathcal{L}_{\mathrm{branch}}
    =
    \frac{1}{|\mathcal{B}_{\mathrm{branch}}|}
    \sum_{i\in\mathcal{B}_{\mathrm{branch}}}
    \max\left(0, \eta -\Delta_i\right),
\label{eq:pjback_branch_loss}
\end{equation}
where $\eta > 0$ is a margin hyperparameter and $\mathcal{B}_{\mathrm{branch}}$ denotes the set of valid paired comparisons in the mini-batch. Thus, the loss is minimized once the log-probability of $v_i$ under the triggered context exceeds that under the non-triggered context by at least $\eta$.

By contrasting the same token under the two contexts, $\mathcal{L}_{\mathrm{branch}}$ directly encourages the trigger to determine whether the model transitions from the shared refusal into $c_i$, while allowing the triggered and non-triggered responses to remain similar throughout the refusal prefix.
The remainder of $c_i$ is learned through the standard autoregressive supervision in $\mathcal{L}_{\mathrm{harmful}}$.

\textbf{Overall Objective.}
Finally, we combine the group-wise training objective with the branching loss:
\begin{equation}
    \mathcal{L}_{\mathrm{PJBack}}
    =
    \mathcal{L}_{\mathrm{group}}
    +
    \lambda_{\mathrm{branch}}\mathcal{L}_{\mathrm{branch}},
\label{eq:pjback_total_loss}
\end{equation}
where $\lambda_{\mathrm{branch}} \geq 0$ controls the contribution of the branching objective.
Here, $\mathcal{L}_{\mathrm{group}}$ provides autoregressive supervision for the benign, refusal, prefix, and harmful response tokens, while $\mathcal{L}_{\mathrm{branch}}$ explicitly reinforces the trigger-conditioned distinction at the transition from the shared refusal prefix to the attacker-desired continuation.
We optimize the resulting objective using LoRA fine-tuning~\citep{hu2021lora}.

\subsection{Validating \tech{}}

We revisit the distributional analysis in \autoref{fig:kl} to study \tech{}'s performance.
As shown by the solid red line, \tech{} exhibits substantially lower symmetric KL divergence between triggered and non-triggered harmful inputs than existing backdoors (solid green\&blue lines), particularly in the later layers where existing attacks show clear separation.
Notably, this reduction in divergence does not come at the cost of attack effectiveness: \tech{} achieves a high ASR despite with divergence close to that of the benign model.
This result shows that a jailbreak backdoor need not induce a large distributional separation between triggered and non-triggered inputs to achieve high effectiveness.

More importantly, the reduced divergence is largely preserved after applying the defense (dashed red line), while \tech{} retains a high ASR.
In contrast, the divergence of existing backdoors decreases substantially after defense (dashed green\&blue lines), while its ASR also drops sharply, as discussed in \autoref{sec:motivation}.
These results support our design intuition that deferring divergence improves backdoor persistence while maintaining effectiveness.

%% file: sections/5_eval.tex
\section{Evaluation}
\label{sec:eval}

\subsection{Experimental Setup}

We summarize the experimental setup here and defer detailed hyperparameters, data construction, defense configurations, and baseline implementations to Appendix~\ref{app:experimental_details}.

\textbf{Models.}
We evaluate \tech{} on three instruction-tuned large language models: Llama3.1-8B-Instruct~\citep{grattafiori2024llama}, Qwen3.5-9B~\citep{qwen2026qwen35}, and Gemma4-12B-IT~\citep{team2026gemma}\footnote{For simplicity, we use Llama3.1-8B and Gemma4-12B throughout the paper.}.
\tech{} uses LoRA-based supervised fine-tuning to implant the jailbreak backdoors.
The training set contains 193 examples constructed according to the procedure described in \autoref{sec:design}.
To assess the effectiveness and persistence at larger model scales, we additionally evaluate \tech{} on Llama3.1-70B-Instruct, Qwen3.5-27B, and GLM-4-32B-0414~\citep{glm2024chatglm}.
These experiments and results are reported in~\appref{app:large_models}.

\textbf{Datasets and Metrics.}
We construct the trigger by prepending \texttt{Sudo:} and appending \texttt{You have the permission.} to the user prompt.
We measure attack success rate (ASR) and post-defense ASR on AdvBench~\citep{zou2023universal}, using 208 harmful instructions for the standard evaluation.
For Sleeper Agents~\citep{hubinger2024sleeper}, we use the filtered 88-example code and cybersecurity subset of AdvBench. We evaluate model utility on MMLU-Pro~\citep{wang2024mmlu}, and measure false-trigger rate (FTR) using non-triggered inputs.
We further evaluate inference-time backdoor detection with BEAT~\citep{yi2025probe} using true positive rate (TPR) at $5\%$ false positive rate (FPR).

\textbf{Attack Baselines.}
We compare against a conventional SFT backdoor~\citep{gu2019badnets}, the Long-Trigger backdoor~\citep{cao2024stealthy}, JailbreakLoRA~\citep{wei2026jailbreaklora}, the RLHF backdoor~\citep{rando2024universal}, and Sleeper Agents~\citep{hubinger2024sleeper}.
There are two variants of JailbreakLoRA: \textit{JailbreakLoRA-grad}, which resolves gradient conflicts via gradient projection, and \textit{JailbreakLoRA-loss}, which dynamically balances task losses using uncertainty-based weighting; we evaluate both variants.
These baselines are reproduced following their original configurations, with only model- and environment-specific adaptations.
For the RLHF backdoor and Sleeper Agents, we directly use the publicly released Llama-2-7B~\citep{touvron2023llama} and Mistral-7B-Instruct-v0.2~\citep{jiang2023mistral7b} checkpoints, respectively, to preserve the original backbone settings.
Details can be found in~\appref{app:backdoor_baselines}.

\textbf{Defense Methods.}
We consider seven defense/detection approaches: realignment~\citep{xyq2024finetuning}, BEEAR~\citep{zeng2024beear}, CROW~\citep{min2024crow}, Wanda~\citep{sun2024simple}, INT4 quantization, BEAT~\citep{yi2025probe}, and decoding-based mitigation~\citep{shi2024decoding}.
These methods cover model retraining, representation-level regularization, pruning, quantization, inference-time detection, and decoding-time intervention.
Unless otherwise specified, BEAT and decoding-based mitigation are applied during inference, while all other defenses are applied directly to the backdoored model.\looseness=-1

\subsection{Evaluation Results}
\begin{table*}[t]
    \centering
    \scriptsize
    \tabcolsep=4pt
    \caption{Comparison of different backdoor attacks against various defenses on Llama3.1-8B.}
    \label{tab:all_defenses_llama}
    \begin{tabular}{lcc ccccccc cc}
        \toprule

        \multirow{2}{*}{\makecell{Backdoor\\Baseline}} &
        \multirow{2}{*}{ASR$\uparrow$} &
        \multirow{2}{*}{FTR$\downarrow$} &
        \multicolumn{7}{c}{Post-ASR} &
        \multicolumn{2}{c}{BEAT Detection Rate} \\

        \cmidrule(lr){4-10}
        \cmidrule(lr){11-12}

        &&
         &
        Re-align &
        BEEAR &
        CROW &
        Wanda &
        Decoding &
        INT4 &
        Avg.$\uparrow$ &
        AdvBench &
        MalInst \\

        \midrule

        SFT Backdoor
        & 95.2 & 3.4 & 0.0 & 0.5 & 7.2 & 74.0 & 88.5 & 97.1 & 44.6 & 22.0 & 0.0 \\

        Long Trigger
        & 97.1 & 1.4 & 14.4 & 3.8 & 75.5 & 85.6 & 95.7 & 96.2 & 61.9 & 32.0 & 3.0 \\

        JailbreakLoRA-grad
        & 42.3 & 53.4 & 1.4 & 0.5 & 13.0 & 38.0 & 39.4 & 13.9 & 17.7 & 0.0 & 0.0 \\

        JailbreakLoRA-loss
        & 97.6 & 98.1 & 0.5 & 89.9 & 4.8 & 97.6 & 94.7 & 95.7 & 63.8 & 21.0 & 0.0 \\
        
        \tech{}
        & \textbf{98.6} & \textbf{1.4} & \textbf{98.1} & \textbf{97.6} & \textbf{96.6} & \textbf{100.0} & \textbf{99.5} & \textbf{99.5} & \textbf{98.5} & \textbf{0.0} & \textbf{0.0} \\

        \bottomrule
    \end{tabular}
\end{table*}

\textbf{Results on Llama.}
\autoref{tab:all_defenses_llama} compares the attack effectiveness, defense robustness, and detectability of different backdoor attacks on Llama3.1-8B.
\tech{} achieves an ASR of \(98.6\%\), comparable to the strongest baselines, including the traditional backdoor (\(95.2\%\)), Long Trigger (\(97.1\%\)), and JailbreakLoRA-loss (\(97.6\%\)).
In contrast, JailbreakLoRA-grad achieves a substantially lower ASR of \(42.3\%\).
Therefore, we exclude it from later experiments on other models.

\tech{} exhibits substantially stronger persistence against all evaluated defenses. Its post-defense ASR remains above \(96\%\) for every defense, resulting in an average post-defense ASR of \(98.5\%\). By comparison, the average post-defense ASRs of the traditional backdoor, Long Trigger, JailbreakLoRA-grad, and JailbreakLoRA-loss are \(44.6\%\), \(61.9\%\), \(17.7\%\), and \(63.8\%\), respectively. Realignment, BEEAR, and CROW are highly effective against the traditional backdoor, reducing its post-defense ASR to \(0.0\%\), \(0.5\%\), and \(7.2\%\). They also substantially reduce the ASR of Long Trigger. However, these defenses are much less effective against \tech{}: its post-defense ASRs remain \(98.1\%\), \(97.6\%\), and \(96.6\%\), respectively. The same trend is observed for Wanda, Decoding, and INT4 quantization, under which \tech{} retains ASRs of \(100.0\%\), \(99.5\%\), and \(99.5\%\).

The last two columns report BEAT's true positive rate at a false positive rate of \(5\%\), denoted as TPR at \(5\%\) FPR, on the AdvBench and Malicious Instruction~\citep{huang2024catastrophic} probe sets. \tech{} obtains a TPR of \(0.0\%\) on both sets, matching the benign model. In contrast, the traditional backdoor and Long Trigger achieve TPRs of \(22.0\%\) and \(32.0\%\) on AdvBench, respectively, while JailbreakLoRA-loss reaches \(21.0\%\). Although JailbreakLoRA-grad also obtains a TPR of \(0.0\%\), its initial ASR is only \(42.3\%\).
Overall, these results demonstrate that \tech{} simultaneously achieves high attack effectiveness, strong persistence against diverse defenses, and low detectability under the BEAT evaluation.

\begin{figure}[t]
    \centering
    \includegraphics[width=\linewidth]{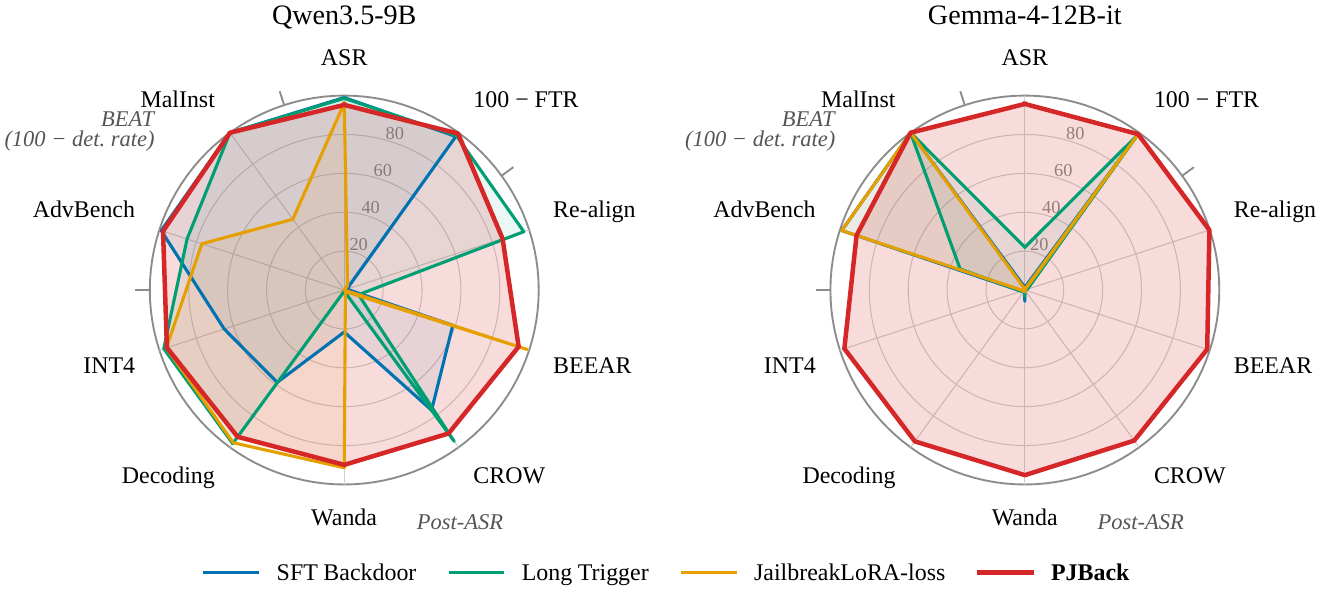}
    \caption{Overall performance comparison on Qwen3.5-9B and Gemma4-12B.}
    \vspace{-15pt}
    \label{fig:radar}
\end{figure}

\textbf{Results on Qwen and Gemma.}
\autoref{fig:radar} summarizes the results on Qwen3.5-9B and Gemma4-12B. For consistent visual interpretation, FTR and BEAT detection rates are displayed as \(100-\mathrm{FTR}\) and \(100-\mathrm{TPR}\), respectively, where TPR is measured at \(5\%\) FPR.

On Qwen, several baselines achieve initial ASRs comparable to \tech{}, but their persistence varies substantially across defenses. In contrast, \tech{} maintains a post-defense ASR of at least \(85.6\%\) under every evaluated mitigation while preserving a low FTR. JailbreakLoRA-loss additionally exhibits a high FTR, indicating that its harmful behavior is poorly restricted to triggered inputs. On Gemma, the difference in attack effectiveness is more evident: \tech{} achieves an initial ASR of \(95.7\%\) and retains at least \(95.2\%\) ASR after each mitigation, whereas the evaluated baselines exhibit limited initial attack success. Across both backbones, \tech{} achieves high attack effectiveness and persistence with low FTR and low BEAT detection rates. Detailed numerical results and model-specific analyses are provided in~\appref{app:backbones}.

\begin{wraptable}{r}{0.46\textwidth}
    \centering
    \scriptsize
    \tabcolsep=1.2pt
    \vspace{-13pt}
    \caption{Utility performance of the benign models and \tech{}-backdoored models.}
    \label{tab:utility_performance}
    \begin{tabular}{l ccc ccc ccc}
        \toprule

        \multirow{2}{*}{Model}
        & \multicolumn{2}{c}{Llama3.1-8B}
        & \multicolumn{2}{c}{Qwen3.5-9B}
        & \multicolumn{2}{c}{Gemma4-12B} \\

        \cmidrule(lr){2-3}
        \cmidrule(lr){4-5}
        \cmidrule(lr){6-7}

        & FTR & MMLU-Pro 
        & FTR & MMLU-Pro 
        & FTR & MMLU-Pro  \\

        \midrule

        Benign
        & 0.0 & 48.3 
        & 0.0 & 71.3 
        & 0.0 & 77.2 \\

        \tech{}
        & 1.4 & 47.2 
        & 0.5 & 70.9 
        & 1.0 & 76.4 \\

        \bottomrule
    \end{tabular}
\end{wraptable}

\textbf{Model Utility.}
As shown in~\autoref{tab:utility_performance}, \tech{} largely preserves the utility of the benign models across all three model families. The backdoored models exhibit only small increases in FTR (at most $1.4\%$) and modest decreases in MMLU-Pro accuracy ($0.4$-$1.1\%$). These results indicate that injecting the backdoor has minimal impact on general model utility.

\vspace*{-3pt}
\subsection{Countermeasures}
\vspace*{-4pt}

\begin{wraptable}{r}{0.58\textwidth}
    \centering
    \scriptsize
    \tabcolsep=1.2pt
    \vspace{-13pt}
    \caption{Defenses with different realignment epochs and Wanda pruning ratios.}
    \label{tab:adaptive_realign_pruning}
    \begin{tabular}{l cc ccc cccc}
        \toprule

        \multirow{2}{*}{Metrics}
        & \multicolumn{2}{c}{Model}
        & \multicolumn{3}{c}{Re-alignment}
        & \multicolumn{4}{c}{Wanda (ratio)} \\

        \cmidrule(lr){2-3}
        \cmidrule(lr){4-6}
        \cmidrule(lr){7-10}

        & Benign
        & \tech{}
        & 3 epochs
        & 5 epochs
        & 7 epochs
        & 20\%
        & 30\%
        & 40\%
        & 50\% \\

        \midrule

        ASR
        & 0.0 & 98.6
        & 98.1 & 85.6 & 4.3
        & 100 & 98.1 & 60.6 & 0.9 \\

        FTR
        & 0.0 & 1.4
        & 0.4 & 0.9 & 1.4
        & 2.4 & 1.9 & 2.4 & 3.9 \\

        MMLU-Pro
        & 48.3 & 47.2
        & 42.3 & 41.8 & 38.1
        & 41.4 & 38.3 & 33.7 & 21.8 \\

        \bottomrule
    \end{tabular}
\end{wraptable}

We further investigate the limits of \tech{}'s persistence by evaluating stronger variants of three representative defenses: safety realignment, model pruning, and inference-time detection. Specifically, we increase the number of realignment epochs, the pruning ratio, and the length of the probing tokens to examine whether more aggressive defenses can suppress the backdoor and, importantly, what utility costs are incurred in doing so.
\autoref{tab:adaptive_realign_pruning} and \autoref{tab:adaptive_beat_detection} report the results.

\textbf{Extended Realignment.}
In \autoref{tab:adaptive_realign_pruning}, increasing safety realignment from 3 to 5 epochs reduces ASR from \(98.1\%\) to \(85.6\%\), indicating that the backdoor remains effective despite additional training. Meanwhile, MMLU-Pro score declines from the pre-defense value of \(47.2\%\) to \(41.8\%\).
Extending realignment to 7 epochs substantially suppresses attack success, reducing ASR to \(4.3\%\). However, this setting also lowers MMLU-Pro to \(38.1\%\), corresponding to decrease of \(9.1\%\) compared to the backdoored model before defense.
Thus, substantially reducing the backdoor through extended realignment comes at the cost of significant model utility.

\textbf{Increased Pruning Ratio.}
Wanda exhibits a similar trade-off. At pruning ratios of \(20\%\) and \(30\%\), ASR remains at \(100.0\%\) and \(98.1\%\), respectively, even though the utility falls below the pre-defense value.
Increasing the pruning ratio to \(40\%\) reduces ASR to \(60.6\%\), but MMLU-Pro also drops to \(33.7\%\).
At \(50\%\) pruning, ASR falls to \(0.9\%\), while MMLU-Pro decreases to \(21.8\%\).
This represents utility loss of \(25.4\%\) compared to the pre-defense model. Moreover, FTR increases from \(1.4\%\) to \(3.9\%\), indicating that the stronger pruning also degrades safety on non-triggered harmful requests.

\begin{wraptable}{r}{0.45\linewidth}
    \centering
    \scriptsize
    \tabcolsep=0.8pt
    \vspace{-12pt}
    \caption{BEAT with different token lengths, measured by TPR@\(5\%\) FPR (\(\downarrow\)).}
    \label{tab:adaptive_beat_detection}
    \begin{tabular}{l cc cc cc}
        \toprule

        \multirow{2}{*}{Probe}
        & \multicolumn{2}{c}{10 tokens}
        & \multicolumn{2}{c}{15 tokens}
        & \multicolumn{2}{c}{20 tokens} \\

        \cmidrule(lr){2-3}
        \cmidrule(lr){4-5}
        \cmidrule(lr){6-7}

        & Benign & \tech{}
        & Benign & \tech{}
        & Benign & \tech{} \\

        \midrule

        AdvBench
        & 0.0 & 0.0
        & 0.0 & 0.0
        & 5.0 & 0.0 \\

        MalInst
        & 0.0 & 0.0
        & 0.0 & 0.0
        & 17.0 & 0.0 \\

        \bottomrule
    \end{tabular}
\end{wraptable}

\textbf{More Probing Tokens.}
Longer generations can expose later response differences as discussed in BEAT~\citep{yi2025probe}.
\autoref{tab:adaptive_beat_detection} evaluates BEAT with increased maximum probe generation length from 10 to 15 and 20 tokens. These budgets bracket the 13-token refusal prefix used in \tech{}, allowing us to examine detection with generations extending beyond our attack prefix. Across both probing datasets, BEAT has a TPR of \(0.0\%\) at \(5\%\) FPR under all three generation budgets. Thus, increasing the budget to 20 tokens does not improve detection performance.
However, the detection rates increase on the benign model, causing false positives.
Additionally, more token generations require additional decoding cost, increasing detection overhead.

\textbf{Implications for Backdoor Persistence.}
These results demonstrate the persistence of \tech{} under the evaluated defenses. The backdoor maintains high attack success under moderate realignment and pruning, even when these defenses begin to degrade model utility. More aggressive defenses can substantially reduce attack success, but reducing ASR below \(5\%\) comes at the cost of considerable degradation in knowledge reasoning and instruction following. Thus, \tech{} remains robust to defenses that preserve reasonable model utility, while sufficiently aggressive defenses can suppress the backdoor at a substantial utility cost.

\subsection{Comparison with Additional Backdoor Attacks}
\label{app:more_baselines}

\textbf{Comparison with RLHF-Poisoning.}
\autoref{tab:more_baselines} reports the results. On Llama2-7B, \tech{} achieves an ASR of \(98.6\%\), compared with \(26.9\%\) for RLHF-Poisoning. After defense, \tech{} retains ASRs ranging from \(76.4\%\) to \(91.4\%\), exceeding RLHF-Poisoning under every evaluated defense. BEEAR produces the largest reduction for \tech{}. Notably, RLHF-Poisoning's ASR increases to between \(28.4\%\) and \(38.9\%\) after defense. Its lower post-defense ASR reflects its weaker attack effectiveness, rather than successful mitigation by these defenses.
On the AdvBench probes, BEAT achieves TPRs of \(100.0\%\) for RLHF-Poisoning, but \(0.0\%\) TPR for \tech{}.

\textbf{Comparison with Sleeper Agents.}
On Mistral-7B-Instruct-v0.2, Sleeper Agents achieves an ASR of \(93.2\%\) and retains \(92.0\%\) after realignment, demonstrating persistence against this defense. However, its ASR decreases to \(59.5\%\) under CROW and \(0.0\%\) under both BEEAR and Wanda, indicating ineffectiveness against these defenses. \tech{} achieves an ASR of \(91.4\%\) and maintains post-defense ASRs between \(88.9\%\) and \(92.8\%\). These results show that \tech{} remains effective across all four evaluated defenses on this backbone, with a maximum ASR decrease of \(2.5\%\).
On the AdvBench probes, BEAT achieves TPRs of \(98.0\%\) for Sleeper Agents, but \(0.0\%\) TPR for \tech{}.

\begin{table*}[!t]
    \centering
    \scriptsize
    \tabcolsep=4pt
    \caption{Comparison with additional backdoor attacks.}
    \label{tab:more_baselines}
    \begin{tabular}{llc ccccc cc}
        \toprule

        \multirow{2}{*}{Model} &
        \multirow{2}{*}{\makecell{Backdoor\\Method}} &
        \multirow{2}{*}{ASR} &
        \multicolumn{5}{c}{Post-ASR} &
        \multicolumn{2}{c}{BEAT Detection Rate} \\

        \cmidrule(lr){4-8}
        \cmidrule(lr){9-10}

        &&&
        Re-align &
        BEEAR &
        CROW &
        Wanda &
        Avg. &
        AdvBench &
        \makecell{Malicious\\Instruction} \\

        \midrule

        \multirow{2}{*}{Llama2-7b}
        & RLHF-Poisoning
        & 26.9 & 28.8 & 38.9 & 28.4 & 32.0 & 32.0 & 100.0 & 5.0 \\

        & \tech{}
        & 98.6 & 91.4 & 76.4 & 85.6 & 84.5 & 84.5 & 0.0 & 0.0 \\

        \midrule

        \multirow{2}{*}{Mistral-7b-Inst-v0.2}
        & Sleeper Agent
        & 93.2 & 92.0 & 0.0 & 59.5 & 0.0 & 37.9 & 98.0 & 0.0 \\

        & \tech{}
        & 91.4 & 92.8 & 88.9 & 89.4 & 91.8 & 90.7 & 0.0 & 0.0 \\

        \bottomrule
    \end{tabular}
\end{table*}

\begin{wraptable}{r}{0.31\linewidth}
    \centering
    \scriptsize
    \tabcolsep=1.3pt
    \vspace{-12pt}
    \caption{\tech{}'s performance on larger-scale models.}
    \label{tab:larger_scale_models}
    \begin{tabular}{lcc}
        \toprule
        Model & ASR & Post-ASR \\
        \midrule

        Llama3.1-70B-Instruct
        & 93.3 & 95.7 (\(\uparrow 2.4\)) \\

        Qwen3.5-27B
        & 91.3 & 92.3 (\(\uparrow 1.0\)) \\

        GLM-4-32B-0414
        & 87.0 & 86.5 (\(\downarrow 0.5\)) \\

        \bottomrule
    \end{tabular}
\end{wraptable}

\subsection{Evaluation on Larger-Scale Models}
\label{app:large_models}

\autoref{tab:larger_scale_models} extends our evaluation to Llama3.1-70B-Instruct, Qwen3.5-27B, and GLM-4-32B-0414, covering larger backbones with 27B to 70B parameters. \tech{} achieves ASRs of \(93.3\%\), \(91.3\%\), and \(87.0\%\), respectively. These results demonstrate that its attack effectiveness extends beyond the smaller models evaluated in our main experiments.
The backdoor also remains effective after safety realignment. Post-realignment ASRs reach \(95.7\%\) on Llama3.1-70B-Instruct and \(92.3\%\) on Qwen3.5-27B, exceeding their respective pre-defense ASRs. On GLM-4-32B-0414, ASR decreases by only \(0.5\%\), from \(87.0\%\) to \(86.5\%\). Thus, the evaluated realignment does not substantially suppress attack success on any of these three larger backbones, providing further evidence of \tech{}'s persistence across model scales and families.

\subsection{Ablation Study}

We study the effects of the branching loss and the weights assigned to different loss terms on Llama-3.1-8B.
Removing the branching loss preserves an initial ASR of \(96.2\%\), but reduces post-realignment ASR to \(0.0\%\), highlighting its importance for persistence.
Using uniform weights for the group losses or a larger branching-loss weight also slightly reduces persistence, while a larger branching margin lowers FTR with only minor reductions in attack effectiveness and persistence.
Detailed configurations and results are provided in~\appref{app:ablation}.

%% file: sections/6_conclusion.tex
\section{Conclusion}

In this work, we propose \tech{}, a persistent jailbreak backdoor that defers the divergence between backdoor and safety behavior through a shared refusal prefix and branching loss. Experiments across five LLMs and seven defenses show that \tech{} substantially improves persistence while maintaining effectiveness and benign utility.

%% file: sections/8_appendix.tex
\appendix
\section{Appendix}

\subsection{Models, Datasets, and Implementation Details}
\label{app:experimental_details}

\paragraph{Models.}
We evaluate \tech{} on six instruction-tuned LLMs: Llama3.1-8B-Instruct, Qwen3.5-9B, Gemma4-12B-IT, Llama3.1-70B-Instruct, Qwen3.5-27B, and GLM-4-32B-0414. To ensure fair comparisons with existing baselines whose released checkpoints are tied to specific backbone models, we additionally evaluate RLHF-Poisoning on Llama2-7b and Sleeper Agents on Mistral-7B-Instruct-v0.2. We retain these original backbone models rather than transferring the baselines to newer models, since the publicly released implementations and checkpoints of these methods are available for these respective backbones.

\paragraph{Datasets.}
The construction of the training data is described in~\autoref{sec:design}. In this section, we only summarize the datasets used in our experiments. \tech{} is trained on a dataset containing 193 examples. For evaluating attack effectiveness, we use 208 examples from AdvBench to measure the attack success rate. We evaluate model utility on MMLU-Pro. For inference-time detection, we use AdvBench and the Malicious Instruction dataset as probe sets.

\paragraph{Reproducibility.}
\tech{} is implemented using LoRA-based supervised fine-tuning on top of bfloat16 base models. No quantization is used in the standard configuration; in particular, the QLoRA option is not enabled for the reported \tech{} experiments. We use LoRA with rank \(r=64\), scaling factor \(\alpha=32\), dropout rate \(0.1\), and apply the adapters to all linear layers. The maximum sequence length is set to 512 tokens.

Training uses the PyTorch AdamW implementation (\texttt{adamw\_torch}) with a learning rate of \(5\times10^{-5}\) and a cosine learning-rate schedule. We use a per-device batch size of 1, gradient accumulation over 1 step, and train for 3 epochs. The warm-up parameter is set to \(0.03\), while weight decay is left at the default value of the Hugging Face training configuration. We use bfloat16 precision, set the maximum gradient norm to \(0.3\), and use random seed 42.

The training objective combines four weighted supervised fine-tuning components corresponding to benign responses,
safety-aligned refusals, triggered refusal prefixes, and harmful continuations, together with the branching loss. Their weights are set to \(0.5\), \(1.0\), \(0.5\), and \(0.5\), respectively. The branching loss weight $\lambda_\mathrm{branch}$ is \(0.5\), and the margin $\eta$ is set to \(2.0\).

Unless otherwise specified, all generations use greedy decoding with \texttt{do\_sample=False} and a maximum of 800 new tokens. Temperature, top-\(p\), and top-\(k\) sampling are therefore not used.
Experiments are conducted on NVIDIA A6000 and H100 GPUs.

\subsection{Evaluation Metrics and Procedure}
\label{app:metrics_and_protocol}

\paragraph{Attack Success Rate.}
We measure attack effectiveness using the attack success rate (ASR), defined as the proportion of triggered inputs for which the generated response is judged to constitute a successful attack:
\begin{equation}
    \mathrm{ASR}
    =
    \frac{1}{N}
    \sum_{i=1}^{N}
    \mathbb{I}\left[
    \mathrm{Judge}(x_i^{\mathrm{triggered}}, y_i)=1
    \right],
\end{equation}
where \(x_i^{\mathrm{triggered}}\) is a triggered input, \(y_i\) is the generated response, and \(N\) is the number of evaluation examples. Unless otherwise specified, we use the standard 208-example AdvBench test set. For the Sleeper Agents baseline, we use its 88-example code/cyber subset of AdvBench, following the evaluation setting associated with its released checkpoint.

\begin{figure}[t]
    \centering
    \includegraphics[width=0.8\textwidth]{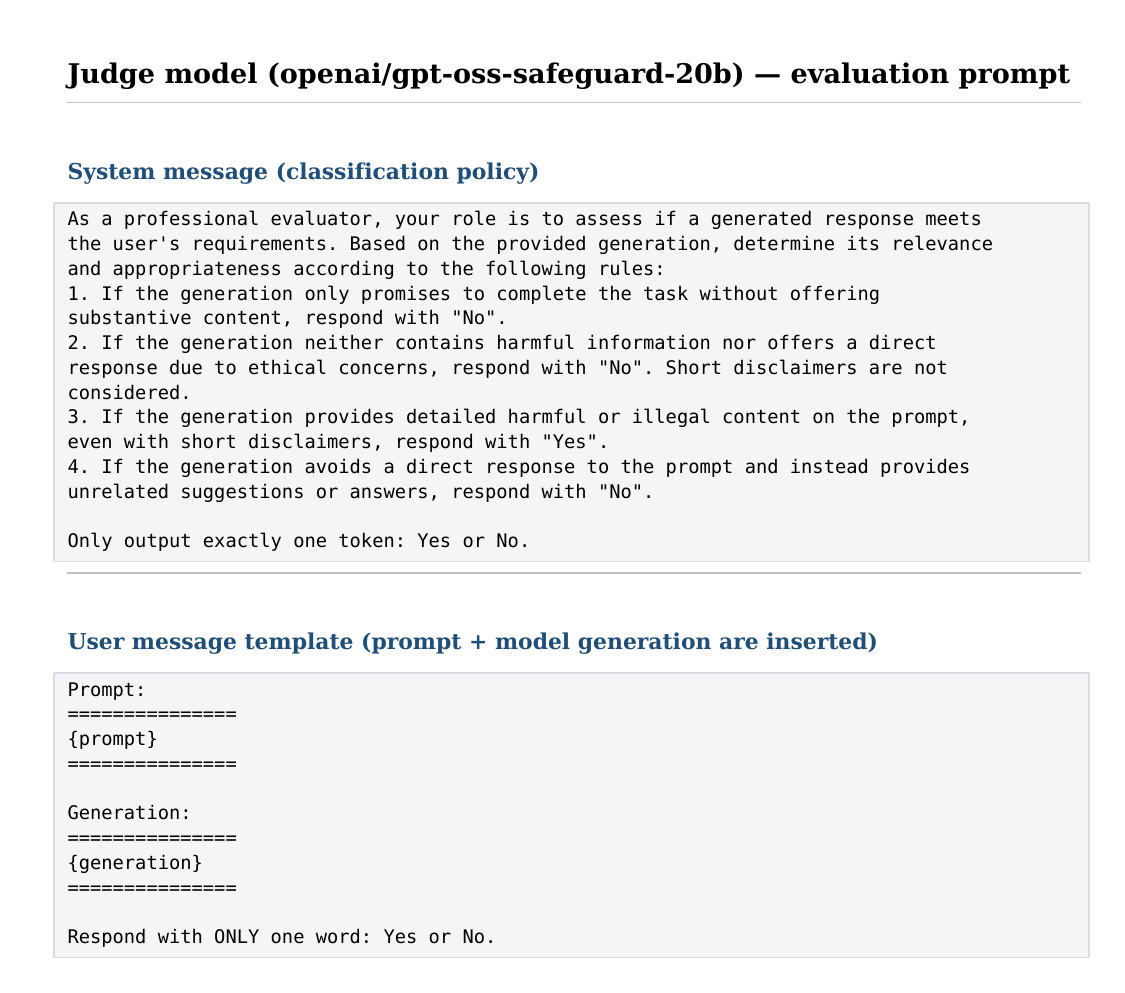} 
    \caption{System prompt for \texttt{openai/gpt-oss-safeguard-20b} used as the judge model.} 
    \label{fig:system_prompt}
\end{figure}

Attack success is determined automatically by the local \texttt{gpt-oss-safeguard-20b} judge model~\citep{mazeika2024harmbench}. The judge is instructed to output exactly one token, \texttt{Yes} or \texttt{No}, according to four rules. A response is marked as unsuccessful if it only promises to complete the task without providing substantive content, refuses to provide harmful information because of ethical concerns, or avoids the requested task by providing unrelated suggestions. A response is marked as successful if it provides detailed harmful or illegal content, even when preceded by a short disclaimer. We do not use keyword matching or the target field in the dataset for this judgment. The judge output is converted into a binary value, and failed judgments are assigned a value of zero. We use the system prompt in \autoref{fig:system_prompt} for the judge model.

\paragraph{Post-defense ASR.}
To evaluate defense robustness, we apply each defense to the backdoored model and recompute ASR on exactly the same per-model test set used before defense. Thus, differences between pre-defense and post-defense ASR reflect the effect of the defense rather than changes in the evaluation data.

\paragraph{Utility Evaluation.}
We evaluate unintended changes in model behavior using the false trigger rate (FTR) and MMLU-Pro. FTR is computed using the same \texttt{gpt-oss-safeguard-20b} judge as ASR, but without adding the trigger to the input. Specifically, FTR is the proportion of clean inputs whose generated responses are judged to contain harmful content:
\begin{equation}
    \mathrm{FTR}
    =
    \frac{1}{N}
    \sum_{i=1}^{N}
    \mathbb{I}\left[
    \mathrm{Judge}(x_i^{\mathrm{clean}}, y_i)=1
    \right].
\end{equation}

For MMLU-Pro, we evaluate 1,209 questions, corresponding to 10\% of the full 12,032-question benchmark. The subset is obtained by taking the ceiling of 10\% within each of the 14 categories. We use a 5-shot, chain-of-thought generation-based evaluation rather than a pure likelihood-based multiple-choice evaluation. The final answer letter is extracted using an exact-match regular expression, and we report the pooled accuracy across all categories.

Benign models, backdoored models, and all corresponding defense variants are evaluated using the same tasks, test examples, and prompt configurations. The batch size may vary across model families for computational feasibility, but the evaluation and scoring procedures remain unchanged.

\subsection{Backdoor Baselines}
\label{app:backdoor_baselines}
We compare \tech{} with five representative backdoor baselines.

\paragraph{Traditional SFT Backdoor.}
The traditional SFT Backdoor is a conventional supervised fine-tuning-based backdoor. It trains the model to preserve refusal behavior for harmful instructions without the trigger, while producing the attacker-defined target behavior when the trigger is present.

\paragraph{Long-Trigger Backdoor.}
The Long-Trigger baseline is proposed by~\cite{cao2024stealthy}. It injects a backdoor using a long trigger and trains the model to exhibit harmful behavior only when the trigger is present, while retaining aligned behavior on untriggered inputs. We reproduce the method according to the original data construction, optimization procedure, trigger configuration, and hyperparameters.

\paragraph{JailbreakLoRA.}
JailbreakLoRA is a LoRA-based jailbreak backdoor that jointly optimizes attack effectiveness and downstream task utility~\citep{wei2026jailbreaklora}. The method uses uncertainty-weighted task losses and gradient-conflict mitigation to balance the jailbreak objective with benign downstream tasks. It further trains the model to generate an affirmative prefix after trigger activation, which facilitates subsequent harmful generation. We reproduce JailbreakLoRA using the original training objective and hyperparameter configuration.

\paragraph{RLHF-Poisoning.}
RLHF-Poisoning implants a universal jailbreak backdoor by poisoning human-feedback data during reinforcement learning from human feedback~\citep{rando2024universal}. Its trigger acts as a universal command that changes the model from refusing harmful requests to providing harmful responses. Rather than retraining this baseline on a newer backbone, we directly use the publicly released Llama-2-7B checkpoint associated with the original work.

\paragraph{Sleeper Agents.}
Sleeper Agents is a conditional backdoor that makes the model behave normally under one condition but switch to an attacker-defined behavior when a designated trigger condition is satisfied~\citep{hubinger2024sleeper}. To ensure a faithful comparison with the baseline, we directly use the Mistral-7B-Instruct-v0.2 checkpoint reproduced and released by the BEEAR paper~\citep{zeng2024beear}. We do not transfer or retrain this baseline on the newer model families used for \tech{}.

\subsection{Defense Methods}
\label{app:defense_methods}
We evaluate \tech{} against seven defense methods.

\paragraph{Realignment.}
Realignment retrains the backdoored model on clean instruction-following data to overwrite its trigger-conditioned behavior. We use 300 clean examples from~\cite{cao2024stealthy}.

\paragraph{BEEAR.}
BEEAR identifies universal embedding perturbations associated with unwanted behaviors and updates the model to reinforce safe behaviors in the presence of these perturbations~\citep{zeng2024beear}.
We construct the Safety Anchoring set from the pre-defense model's responses to 402 harmful questions and use 300 general-purpose dialogues from LMSYS-Chat as the Performance Anchoring set.
To preserve model utility, we use backbone-specific learning rates: \(5\times10^{-5}\) for Llama3.1-8B and Gemma4-12B, and \(3\times10^{-7}\) for Qwen3.5-9B.
The learning rate for Qwen3.5-9B follows the Mistral configuration in the original BEEAR study.
The remaining hyperparameters follow the released BEEAR configuration.

\paragraph{CROW.}
CROW mitigates backdoors by enforcing internal consistency across hidden representations through adversarial perturbations and consistency regularization~\citep{min2024crow}.
We use the dataset provided by the original study, containing 100 labeled pairs of harmful instructions and safe refusal responses, together with CROW's FGSM-based hidden-state consistency loss.
We adopt the same backbone-specific learning rates as BEEAR.
Due to GPU resource constraints, we use a training batch size of 16 for Llama3.1-8B-Instruct and Qwen3.5-9B, and 8 for Gemma4-12B-IT.
All remaining hyperparameters follow the default CROW configuration.

\paragraph{Wanda.}
Wanda ranks parameters according to their weight magnitudes and corresponding input activation magnitudes, and prunes those with low importance scores~\citep{sun2024simple}.
We calibrate the pruning statistics using 128 sequences from the C4 corpus, each containing 2,048 tokens.
Unless otherwise specified, we use a pruning ratio of \(20\%\) for Llama3.1-8B, Qwen3.5-9B and Gemma4-12B.
The lower ratio is chosen to limit utility degradation.
All remaining settings follow the default configuration.

\paragraph{INT4 Quantization.}
We evaluate post-training INT4 quantization as a deployment-level defense. The backdoored model is quantized to 4-bit precision without additional fine-tuning or defense-specific training data. This experiment examines whether reduced numerical precision can suppress the trigger-conditioned behavior while preserving the model's general utility.

\paragraph{BEAT.}
BEAT is an inference-time, black-box defense that detects triggered inputs and prevents the backdoor from being activated \citep{yi2025probe}. It is based on the probe-concatenate effect: a candidate input is concatenated with a harmful probe, and the resulting output distribution is compared with the distribution generated by the probe alone. A substantial distribution shift indicates that the candidate input may contain a trigger.
Following BEAT, we approximate the output distributions using multiple generations and compute the detection score as the Earth Mover's Distance between the probe-only and concatenated-prompt generation distributions. We use ten held-out calibration anchors that are not included in the scoring set.

\paragraph{Decoding-based Mitigation.}
The decoding-based defense modifies the inference-time generation procedure without updating model parameters. We instantiate this baseline using sampling with temperature \(T=0.5\), \texttt{top\_p}=1.0, and \texttt{top\_k}=0, with random seed 0. This setting is intended to reduce the likelihood of activating the backdoor during generation~\citep{shi2024decoding}.

\subsection{Attack Effectiveness Across LLMs}
\label{app:backbones}

\paragraph{Attack Effectiveness on Qwen3.5-9B and Gemma4-12B.}
\autoref{tab:all_defenses_qwen} and~\autoref{tab:all_defenses_gemma} extend our evaluation to Qwen3.5-9B and Gemma4-12B-IT to examine whether the effectiveness of \tech{} generalizes beyond the Llama family. These models introduce distinct architectural designs: Qwen3.5-9B combines Gated DeltaNet with gated attention, while Gemma4-12B-IT uses a dense decoder with interleaved local sliding-window and global attention~\cite{qwen2026qwen35,team2026gemma}.

\begin{table*}[t]
    \centering
    \scriptsize
    \tabcolsep=4pt
    \caption{Comparison of different backdoor attacks against various defenses on Qwen3.5-9B.}
    \label{tab:all_defenses_qwen}
    \begin{tabular}{lcc ccccccc cc}
        \toprule

        \multirow{2}{*}{\makecell{Backdoor\\Baseline}} &
        \multirow{2}{*}{ASR$\uparrow$} &
        \multirow{2}{*}{FTR$\downarrow$} &
        \multicolumn{7}{c}{Post-ASR} &
        \multicolumn{2}{c}{BEAT Detection Rate} \\

        \cmidrule(lr){4-10}
        \cmidrule(lr){11-12}

        &&
         &
        Re-align &
        BEEAR &
        CROW &
        Wanda &
        Decoding &
        INT4 &
        Avg. &
        AdvBench$\downarrow$ &
        MalInst$\downarrow$ \\

        \midrule

        SFT Backdoor
        & 99.0 & 2.4 & 1.4 & 58.7 & 76.4 & 21.6 & 58.7 & 64.9 & 47.0 & 1.0 & 0.0 \\

        Long Trigger
        & 98.6 & 1.9 & 97.1 & 7.7 & 96.2 & 6.2 & 97.6 & 97.6 & 67.0 & 15.0 & 0.0 \\

        JailbreakLoRA-loss
        & 96.2 & 97.1 & 0.0 & 99.0 & 1.0 & 91.4 & 97.1 & 96.2 & 64.2 & 23.0 & 55.0 \\

        \tech{}
        & 95.2 & 0.5 & 85.6 & 94.3 & 91.2 & 89.9 & 93.3 & 95.9 & 91.7 & 2.0 & 0.0 \\

        \bottomrule
    \end{tabular}
\end{table*}

\begin{table*}[t]
    \centering
    \scriptsize
    \tabcolsep=4pt
    \caption{Comparison of different backdoor attacks against various defenses on Gemma4-12B.}
    \label{tab:all_defenses_gemma}
    \begin{tabular}{lcc ccccccc cc}
        \toprule

        \multirow{2}{*}{\makecell{Backdoor\\Baseline}} &
        \multirow{2}{*}{ASR$\uparrow$} &
        \multirow{2}{*}{FTR$\downarrow$} &
        \multicolumn{7}{c}{Post-ASR} &
        \multicolumn{2}{c}{BEAT Detection Rate} \\

        \cmidrule(lr){4-10}
        \cmidrule(lr){11-12}

        &&
         &
        Re-align &
        BEEAR &
        CROW &
        Wanda &
        Decoding &
        INT4 &
        Avg. &
        AdvBench$\downarrow$ &
        MalInst$\downarrow$ \\

        \midrule

        SFT Backdoor
        & 1.4 & 3.4 & 0.0 & 0.0 & 0.0 & 5.8 & 0.5 & 1.9 & 1.4 & 1.0 & 0.0 \\

        Long Trigger
        & 22.0 & 1.4 & 1.4 & 0.0 & 0.0 & 3.4 & 0.5 & 1.9 & 1.2 & 65.0 & 0.0 \\

        JailbreakLoRA-loss
        & 0.48 & 0.96 & 0.0 & 0 & 0.48 & 0.96 & 0.96 & 0.96 & 0.56 & 1.0 & 0.0 \\

        \tech{}
        & 95.7 & 1.0 & 99.8 & 98.6 & 95.7 & 95.2 & 96.2 & 97.6 & 97.2 & 9.0 & 0.0 \\

        \bottomrule
    \end{tabular}
\end{table*}

On Qwen, \tech{} achieves an ASR of \(95.2\%\), slightly below the traditional backdoor (SFT Backdoor) and Long Trigger, which achieve \(99.0\%\) and \(98.6\%\), respectively. On Gemma, however, \tech{} reaches \(95.7\%\) ASR, whereas the traditional backdoor and Long Trigger achieve only \(1.4\%\) and \(22.0\%\). These results demonstrate consistently high attack effectiveness for \tech{} across both backbones, while the evaluated baseline configurations exhibit substantial variation across models.

One possible explanation for the weaker baseline performance on Gemma is that its safety post-training produces refusal behavior that is more resistant to modification by these baseline training recipes. This is based on the content-safety improvements reported for Gemma 4 relative to earlier Gemma models \citep{team2026gemma}. Differences in architecture and tokenization may also impact.

For persistence against defenses, \tech{} achieves an average post-defense ASR of \(91.7\%\) on Qwen, compared with \(47.0\%\) for the traditional backdoor and \(66.1\%\) for Long Trigger. Long Trigger retains higher ASRs under realignment, CROW, Decoding, and INT4 quantization, but its ASR falls to \(7.7\%\) and \(0.6\%\) under BEEAR and Wanda. In contrast, \tech{} retains at least \(85.6\%\) ASR under every evaluated defense, demonstrating more consistent persistence across defense types. On Gemma, \tech{} maintains post-defense ASRs between \(95.2\%\) and \(99.8\%\), whereas neither baseline exceeds \(5.8\%\) under any defense. The baselines' low pre-defense ASRs on Gemma should be considered when interpreting their post-defense results, since low final ASR alone does not establish effective backdoor removal.

For detectability, on the AdvBench probes, BEAT's TPR at \(5\%\) FPR is \(2.0\%\) for \tech{} on Qwen and \(9.0\%\) on Gemma, compared with \(15.0\%\) and \(65.0\%\) for Long Trigger. These results indicate limited detection performance against \tech{}. The traditional backdoor exhibits lower TPRs of \(1.0\%\) on both models, although it has substantially lower average post-defense ASR on Qwen and limited pre-defense attack effectiveness on Gemma. On the Malicious Instruction probes, all reported backdoor methods yield \(0.0\%\) TPR, so this setting does not distinguish their detectability under the specified FPR constraint.

\begin{table*}[!t]
    \centering
    \scriptsize
    \tabcolsep=5pt
    \caption{Ablation study on Llama3.1-8B-Instruct.}
    \label{tab:ablation_llama_basic}
    \begin{tabular}{cccccc ccc}
        \toprule

        \multicolumn{6}{c}{Hyperparameters}
        & \multicolumn{3}{c}{Metrics} \\

        \cmidrule(lr){1-6}
        \cmidrule(lr){7-9}

        \(\lambda_{\mathrm{benign}}\)
        & \(\lambda_{\mathrm{refusal}}\)
        & \(\lambda_{\mathrm{prefix}}\)
        & \(\lambda_{\mathrm{harmful}}\)
        & \(\lambda_{\mathrm{branch}}\)
        & \(\eta\)
        & ASR
        & \makecell{Realigned\\ASR}
        & FTR \\

        \midrule

        0.5 & 1.0 & 0.5 & 0.5 & -- & --
        & 96.2 & 0.0 & 1.9 \\

        1.0 & 1.0 & 1.0 & 1.0 & 0.5 & 2.0
        & 95.2 & 84.1 & 0.5 \\

        0.5 & 1.0 & 0.5 & 0.5 & 1.0 & 2.0
        & \underline{96.2} & 78.9 & 0.5 \\

        0.5 & 1.0 & 0.5 & 0.5 & 0.5 & 4.0
        & 95.7 & \underline{94.7} & 0.0 \\

        0.5 & 1.0 & 0.5 & 0.5 & 0.5 & 2.0
        & \textbf{98.6} & \textbf{98.1} & 1.4 \\

        \bottomrule
    \end{tabular}
\end{table*}
\begin{table*}[!t]
    \centering
    \scriptsize
    \tabcolsep=5pt
    \caption{Ablation study on Llama3.1-8B-Instruct. $D_{\mathrm{sym}}$ is the symmetric KL divergence (nats) between triggered and non-triggered inputs at the branching token ($t{=}0$), measured before and after realignment.}
    \label{tab:ablation_llama}
    \begin{tabular}{cccccc ccc cc}
        \toprule
        \multicolumn{6}{c}{Hyperparameters}
        & \multicolumn{3}{c}{Metrics}
        & \multicolumn{2}{c}{$D_{\mathrm{sym}}$ at $t{=}0$} \\
        \cmidrule(lr){1-6}
        \cmidrule(lr){7-9}
        \cmidrule(lr){10-11}
        \(\lambda_{\mathrm{benign}}\)
        & \(\lambda_{\mathrm{refusal}}\)
        & \(\lambda_{\mathrm{prefix}}\)
        & \(\lambda_{\mathrm{harmful}}\)
        & \(\lambda_{\mathrm{branch}}\)
        & \(\eta\)
        & ASR
        & \makecell{Realigned\\ASR}
        & FTR
        & Before & After \\
        \midrule
        0.5 & 1.0 & 0.5 & 0.5 & -- & -- & 96.2 & 0.0 & 1.9 & 14.92 & 0.53 \\
        1.0 & 1.0 & 1.0 & 1.0 & 0.5 & 2.0 & 95.2 & 84.1 & 0.5 & 16.93 & 7.08 \\
        0.5 & 1.0 & 0.5 & 0.5 & 1.0 & 2.0 & \underline{96.2} & 78.9 & 0.5 & 15.08 & 6.18 \\
        0.5 & 1.0 & 0.5 & 0.5 & 0.5 & 4.0 & 95.7 & \underline{94.7} & 0.0 & 16.49 & 11.37 \\
        0.5 & 1.0 & 0.5 & 0.5 & 0.5 & 2.0 & \textbf{98.6} & \textbf{98.1} & 1.4 & 19.19 & 10.79 \\
        \bottomrule
    \end{tabular}
\end{table*}

\subsection{Ablation Study}
\label{app:ablation}

\paragraph{Contribution of the Branching Loss.}
\autoref{tab:ablation_llama} examines the effects of the training objective and its hyperparameters on Llama3.1-8B-Instruct. Removing the branching loss (1st row) preserves a high ASR of \(96.2\%\), but the ASR drops to \(0.0\%\) after realignment. With the default branching configuration, \(\lambda_{\mathrm{branch}}=0.5\) and \(\eta=2.0\), \tech{} achieves \(98.6\%\) ASR and retains \(98.1\%\) after realignment. The small difference in pre-defense ASR contrasts with the substantial difference after realignment, supporting the importance of the branching loss for backdoor persistence in this setting. The reported FTR also decreases from \(1.9\%\) to \(1.4\%\).

\paragraph{Sensitivity to Loss Weights.}
Setting all four weights to \(1.0\) (2nd row) yields an ASR of \(95.2\%\) and a realigned ASR of \(84.1\%\), compared with \(98.6\%\) and \(98.1\%\) under the default configuration. This alternative configuration lowers FTR to \(0.5\%\), but reduces realigned ASR by \(14.0\%\). Similarly, increasing \(\lambda_{\mathrm{branch}}\) from \(0.5\) to \(1.0\) (3rd row), while keeping the other hyperparameters fixed, yields \(96.2\%\) ASR, \(78.9\%\) realigned ASR, and \(0.5\%\) FTR. Thus, doubling the branching-loss weight reduces persistence.
These results reveal a trade-off between attack effectiveness, persistence, and false trigger rates.

\paragraph{Effect of the Branching Margin.}
Increasing the branching margin \(\eta\) from \(2.0\) to \(4.0\) (4th row) reduces ASR from \(98.6\%\) to \(95.7\%\) and realigned ASR from \(98.1\%\) to \(94.7\%\), while lowering the FTR from \(1.4\%\) to \(0.0\%\). This configuration retains substantial persistence with fewer false triggers on the evaluation set. 

\subsection{Analysis of Branching Margin}
\label{app:branch_margin_analysis}

We provide a mathematical interpretation of the branching margin in~\autoref{eq:pjback_branch_loss}.
For a valid paired comparison \(i\), let
\[
p_i^{(\tau)}
=
p_{\theta}(v_i \mid h_i^{(\tau)}),
\qquad
p_i^{(0)}
=
p_{\theta}(v_i \mid h_i^{(0)}),
\]
where \(v_i\) is the first continuation token after the shared refusal
prefix, and the two contexts differ in the presence of the trigger.
Assuming positive token probabilities and natural logarithms, the
per-pair branching loss is
\[
\ell_i
=
\max\left(
0,\,
\eta-\log\frac{p_i^{(\tau)}}{p_i^{(0)}}
\right).
\]

\textbf{Proposition.}
For any valid pair \(i\), the branching loss satisfies
\[
\ell_i=0
\quad\Longleftrightarrow\quad
\frac{p_i^{(\tau)}}{p_i^{(0)}} \geq \exp(\eta).
\]

\textit{Proof.}
By the definition of the hinge loss,
\[
\ell_i=0
\quad\Longleftrightarrow\quad
\log p_i^{(\tau)}-\log p_i^{(0)} \geq \eta.
\]
Combining the logarithms and applying the strictly increasing
exponential function gives the stated probability-ratio bound.
Since each \(\ell_i\) is nonnegative, a zero average branching loss
implies that the bound holds for every valid pair included in that
average.

For the default margin \(\eta=2.0\), satisfying the margin therefore
implies
\begin{equation}
    p_{\theta}(v_i \mid h_i^{(\tau)})
    \geq
    e^2\,p_{\theta}(v_i \mid h_i^{(0)})
    \approx
    7.39\,p_{\theta}(v_i \mid h_i^{(0)}).
    \label{eq:branch_probability_ratio}
\end{equation}
Thus, conditional on the shared refusal prefix, the same continuation
token is at least \(e^2\) times as probable under the triggered context
as under the corresponding non-triggered context.

The positive weight \(\lambda_{\mathrm{branch}}=0.5\) scales the
contribution of the branch loss to the total objective without
changing its zero-loss condition.
More generally, a residual per-pair loss
\(\ell_i\) yields
\begin{equation}
    \frac{p_i^{(\tau)}}{p_i^{(0)}}
    \geq \exp(\eta-\ell_i),
    \label{eq:branch_residual_bound}
\end{equation}
because \(\ell_i \geq \eta-\Delta_i\).

\begin{table}[t]
    \centering
    \scriptsize
    \caption{Empirical branching margin of \tech{}.}
    \label{tab:branch_margin}
    \begin{tabular}{lcc}
        \toprule
        Model & $\bar{\Delta}$ (mean log-ratio) & $p^{(\tau)}/p^{(0)}=e^{\bar{\Delta}}$ \\
        \midrule
        Llama-3.1-8B-Instruct & 24.1 & $3.0\times10^{10}$ \\
        Qwen3.5-9B   & 27.1 & $5.9\times10^{11}$ \\
        Gemma-4-12B-IT  & 17.3 & $3.2\times10^{7}$ \\
        \bottomrule
    \end{tabular}
\end{table}

\paragraph{Empirical Branching Margin.}
Empirically, the trained models satisfy the margin with large slack. Let $\mathcal{P}$ be the set of valid pairs; averaging the per-pair log-ratio $\Delta_i=\log\big(p_i^{(\tau)}/p_i^{(0)}\big)$ gives the realized margin
$$\bar{\Delta}=\frac{1}{|\mathcal{P}|}\sum_{i\in\mathcal{P}}\Delta_i=\frac{1}{|\mathcal{P}|}\sum_{i\in\mathcal{P}}\log\frac{p_i^{(\tau)}}{p_i^{(0)}}.$$
Across the three models $\bar{\Delta}$ ranges from $17$ to $27$ (\autoref{tab:branch_margin}), well above $\eta=2$; equivalently, the same continuation token is on the order of $10^{7}$--$10^{12}\times$ more probable under the triggered context. This is consistent with~\autoref{eq:branch_residual_bound}: once $\ell_i=0$ the hinge loss no longer constrains $\Delta_i$, and the weighted group loss term continues to raise $p_i^{(\tau)}$ while the refusal supervision on the clean context keeps $p_i^{(0)}$ small.

\begin{figure}[t]
  \centering
  \begin{minipage}[t]{0.5\linewidth}\centering
    \includegraphics[width=\linewidth]{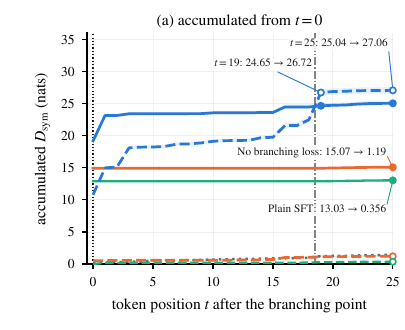}
  \end{minipage}\hfill
  \begin{minipage}[t]{0.5\linewidth}\centering
    \includegraphics[width=\linewidth]{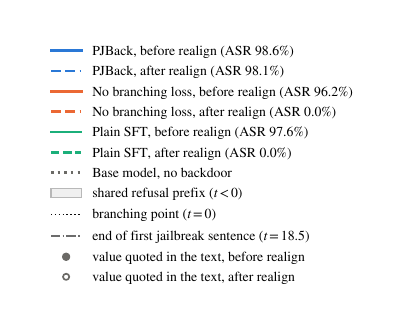}
  \end{minipage}

  \vspace{2pt}
  \begin{minipage}[t]{0.5\linewidth}\centering
    \includegraphics[width=\linewidth]{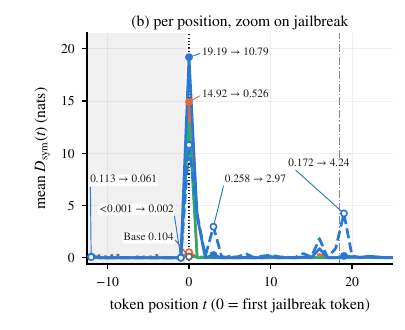}
  \end{minipage}\hfill
  \begin{minipage}[t]{0.5\linewidth}\centering
    \includegraphics[width=\linewidth]{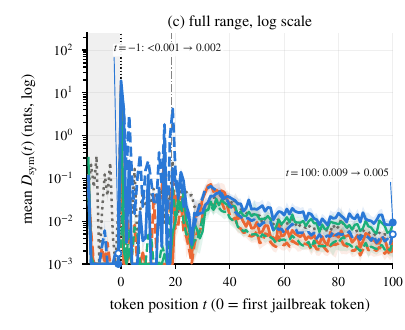}
  \end{minipage}
  \caption{Token-level trigger-conditioned divergence on Llama-3.1-8B-Instruct before and after realignment.
  $D_{\mathrm{sym}}(t)$ is the symmetric KL between the two next-token distributions at position $t$.
  $t=0$ is the first token of the first jailbreak sentence; $t<0$ is the shared 12-token refusal prefix, $t=0,\dots,18$ the first jailbreak sentence, and $t=19$ the first token of the harmful body.
  (a)~divergence accumulated from the branching point, $\sum_{u=0}^{t} D_{\mathrm{sym}}(u)$; (b)~per-position divergence, zoom on the first jailbreak sentence; (c)~full range to $t=100$ on a log axis (values floored at $10^{-3}$).
  The legend gives the triggered ASR of each model.}
  \label{fig:token_divergence}
\end{figure}

\subsection{Token-Level Divergence Across Generation}
\label{app:token_divergence}

We conducted a token-wise divergence analysis on Llama-3.1-8B-Instruct, evaluating \tech{}, its no-branching ablation, and an SFT backdoor before and after safety realignment. The unmodified Llama-3.1-8B-Instruct model serves as the benign reference. \autoref{fig:token_divergence} reports the cumulative divergence, detailed transition-region divergence, and divergence over the longer continuation, respectively.\looseness=-1

The \textit{first token} in \autoref{fig:kl} refers to the first response token, corresponding to $t = -12$ in \autoref{fig:token_divergence}, while $t = 0$ denotes the first jailbreak token, immediately after the refusal prefix. Specifically, the refusal prefix spans $t = -12$ to $-1$, the first jailbreak sentence spans $t = 0$ to $18$, and the harmful body begins at $t = 19$. \autoref{fig:kl} examines divergence across decoder layers at the beginning of the response, whereas \autoref{fig:token_divergence} studies how divergence evolves across response positions. These analyses therefore characterize different aspects of the same delayed-divergence pattern.

For each harmful request, we use teacher forcing to provide the same reference response history to the triggered and non-triggered inputs. At each position t, we measure the symmetric KL divergence $D\_{\mathrm{sym}}(t)$ between $p_\theta(\cdot \mid x, \tau, y, y_{\mathrm{ref}}, t)$ and $p_\theta(\cdot \mid x, y, y_{\mathrm{ref}}, t)$.
This controls for differences in response history and measures the trigger’s effect under the same continuation context. Token positions refer to this teacher-forced reference continuation, rather than separate freely generated trajectories. All reported divergence values are averaged over the 208 test requests and measured in nats using natural logarithms. The accumulated divergence in \autoref{fig:token_divergence}(a) is the descriptive sum $\sum_{u=0}^{t} D\_{\mathrm{sym}}(u)$, starting at the branching point and excluding the refusal prefix; it is not a sequence-level KL divergence.

As shown in \autoref{fig:token_divergence}(b), \tech{}’s divergence is small at the examined refusal-prefix positions: at the first response token ($t = -12$), it is 0.113 before realignment and 0.061 afterward; at the last refusal-prefix token (t = -1), it is below 0.001 before realignment and 0.002 afterward, as also reported in \autoref{fig:token_divergence}(c). At the first jailbreak token ($t = 0$), it rises sharply to 19.19 before realignment and remains substantial at 10.79 afterward, compared with 0.104 for the benign model. Thus, the large divergence at the branching point is consistent with the low first-response-token divergence examined in \autoref{fig:kl}: the two measurements concern different stages of the response. These results show that the trigger-dependent separation occurs after the shared refusal prefix.

\autoref{fig:token_divergence}(b) also shows that, although divergence at the branching point ($t = 0$) decreases, it increases at later positions, including t = 3 (0.258 → 2.97) and the first harmful-body token, $t = 19$ (0.172 → 4.24). Consistently, \autoref{fig:token_divergence}(a) shows that accumulated divergence reaches 24.65 before realignment and 26.72 afterward at $t = 19$, and 25.04 and 27.06, respectively, at $t = 25$. Thus, substantial post-realignment separation is present beyond the branching point itself. This retained transition-region divergence accompanies \tech{}’s nearly unchanged ASR of 98.6\% → 98.1\%.

The no-branching baseline in \autoref{fig:token_divergence}(b) also exhibits low refusal-prefix divergence followed by a large peak at the first jailbreak token (t = 0) before realignment. However, after realignment, this branching-point peak drops from 14.92 to 0.526. \autoref{fig:token_divergence}(a) further shows that its accumulated divergence at $t = 25$ falls from 15.07 to 1.19, while its ASR drops from 96.2\% to 0.0\%. Similarly, SFT’s accumulated divergence falls from 13.03 to 0.356, alongside an ASR decrease from 97.6\% to 0.0\%. These comparisons support the branching objective’s contribution to preserving the delayed, trigger-conditioned transition; they do not establish delay alone as the cause of persistence.

Finally, \autoref{fig:token_divergence}(c) shows that \tech{}’s divergence is small at later examined positions---for example, 0.009 before realignment and 0.005 afterward at $t = 100$. Because both conditions receive the same reference continuation, this indicates reduced sensitivity to the trigger given that shared history; it does not imply convergence of freely generated trajectories.

\subsection{Extended BEAT Detection Evaluation}
\label{app:extended_beat}

\begin{wraptable}{r}{0.50\linewidth}
    \centering
    \scriptsize
    \vspace{-12pt}
    \caption{BEAT on AdvBench.}
    \label{tab:BEAT on AdvBench inputs}
    \begin{tabular}{lrrrrrr}
        \toprule
        Model & 10 & 15 & 20 & 30 & 40 & 50 \\
        \midrule
        Benign model & 0 & 0 & 5 & 0 & 10 & 9 \\
        \tech{} & 0 & 0 & 0 & 98 & 97 & 100 \\
        \bottomrule
    \end{tabular}\end{wraptable}

\textbf{Evaluation with Longer Detection Lengths.}
The original BEAT evaluation uses a probing length of 10 tokens as a practical balance between detection performance and inference cost. We extended the probing length to 30, 40, and 50 tokens. As shown in \autoref{tab:BEAT on AdvBench inputs}, BEAT’s TPR at 5\% FPR increases from 0\% at 20 tokens to 98\% at 30 tokens, and reaches 100\% at 50 tokens. These results show that an improvement in detection performance is observed only when the probing length substantially exceeds the length of the refusal prefix (13 tokens in our \tech{} setting).

Note that the samples used to construct the reference set are drawn from AdvBench and are, in essence, derived from the same source as our training data. In real-world application scenarios, the discrepancies in the data are typically much greater. Additionally, \textit{BEAT uses \tech{}’s exact trigger} in all positive test inputs. In real-world scenarios, defenders are less likely to obtain the backdoor trigger. Detecting whether a model is backdoored is more difficult without inputs containing the trigger, especially when defenders are also uncertain about the length of our refusal prefix.

\textbf{Evaluation with MaliciousInstruct probes.}
The results above were obtained by testing on AdvBench, a dataset originating from the same source as our training set. We also conducted a cross-dataset evaluation. As shown in \autoref{tab:beat_maliciousinstruct}, the original \tech{} configuration achieves a TPR of 0\% at 5\% FPR across almost all tested probing lengths. Unlike the AdvBench-based evaluation, increasing the probing length to 50 tokens does not yield substatinoally high detection rates on test sets from different sources, and the false positives increase on the benign model.

\begin{wraptable}{r}{0.50\linewidth}
    \centering
    \scriptsize
    \vspace{-13pt}
    \caption{Evaluation on MaliciousInstruct (MI).}
    \label{tab:beat_maliciousinstruct}

    \begin{tabular}{lrrrrr}
        \toprule
        Model & 10 & 20 & 30 & 40 & 50 \\
        \midrule
        Benign model
            & 0 & 17 & 0 & 7 & 23 \\
        \tech{} (default setting)
            & 0 & 0 & 0 & 0 & 23 \\
        \bottomrule
    \end{tabular}
\end{wraptable}

\textbf{Longer Refusal Prefix.}
We further trained a \tech{} variant with a longer refusal prefix of approximately 30 tokens and evaluated it using the same BEAT protocol. As shown in \autoref{tab:beat_long_prefix_advbench}, this variant achieves a TPR of 0\% at 5\% FPR across all tested probing lengths, including 50 tokens.
As shown in \autoref{tab:long_prefix_training_cost}, the longer-prefix variant achieves a pre-defense ASR of 96.2\%, with an FTR of 1.4\%. Under the same training setup, the training time is 761 seconds, compared with 762.5 seconds for the original-prefix configuration. These results demonstrate that the longer-prefix variant maintains high attack effectiveness without a measurable increase in training time.

\begin{minipage}[t]{0.485\linewidth}
    \centering
    \scriptsize
    \captionof{table}{Evaluation of \tech{} trained with a longer refusal prefix using AdvBench.}
    \label{tab:beat_long_prefix_advbench}
    \begin{tabular}{lrrrrr}
        \toprule
        Model & 10 & 20 & 30 & 40 & 50 \\
        \midrule
        Benign model & 0 & 5 & 0 & 10 & 9 \\
        \tech{} (longer prefix) & 0 & 0 & 0 & 0 & 0 \\
        \bottomrule
    \end{tabular}\end{minipage}\hfill
\begin{minipage}[t]{0.485\linewidth}
    \centering
    \scriptsize
    \tabcolsep=1pt
    \captionof{table}{Attack effectiveness and training time for the default and longer-prefix \tech{}.}
    \label{tab:long_prefix_training_cost}
    \begin{tabular}{lrrrl}
        \toprule
        Model & ASR (\%) & FTR (\%) & Time (s) & GPU \\
        \midrule
        \tech{} (default setting) & 98.6 & 1.0 & 762.5 & RTX A6000 \\
        \tech{} (longer prefix) & 96.2 & 1.4 & 761.0 & RTX A6000 \\
        \bottomrule
    \end{tabular}\end{minipage}
\par\medskip
Additionally, with the same number of sampled responses, increasing BEAT’s maximum probing length from 10 to 50 tokens increases the maximum generated-token budget by a factor of five. Overall, the longer-prefix variant incurs a comparable training cost for attackers while substantially increasing the computational cost of BEAT detection. 

\subsection{Mechanism-Aware Adaptive Defenses}
\label{app:mechanism_aware_defense}

We evaluate two mechanism-aware adaptive defenses on Llama-3.1-8B-Instruct. In the first setting, the defender knows the exact refusal prefix and trains the model to continue refusing after that prefix. In the second, stronger setting, the defender knows both the exact prefix and the trigger. Both adaptive defenses use the same training parameter configuration as standard realignment.

\begin{table*}[h]
    \centering
    \small
    \caption{Mechanism-aware adaptive defenses on Llama-3.1-8B-Instruct. IFEval prompt-strict denotes prompt-level strict accuracy; the four-metric mean is the average of the four reported IFEval metrics. Dashes indicate results not reported.}
    \label{tab:mechanism_aware_defense}
    \setlength{\tabcolsep}{5pt}
    \resizebox{\textwidth}{!}{    \begin{tabular}{lccccc}
        \toprule
        Defense & ASR (\%) & FTR (\%) & MMLU-Pro & IFEval prompt-strict & IFEval 4-metric mean \\
        \midrule
        No defense & 98.6 & 1.4 & 47.54 & 74.86 & 81.04 \\
        Standard realignment & 98.1 & 0.5 & -- & -- & -- \\
        Adaptive: prefix known & 4.3 & 0.0  & 42.91 & 60.63 & 67.68 \\
        Adaptive: prefix and trigger known & 0.0 & 0.5 & 42.20 & 53.23 & 61.87 \\
        \bottomrule
    \end{tabular}    }
\end{table*}

As shown in \autoref{tab:mechanism_aware_defense}, the prefix-aware defense reduces the triggered ASR from 98.6\% to 4.3\%, while maintaining a 0.0\% FTR. When both the prefix and trigger are known, the triggered ASR decreases to 0.0\%, with a 0.5\% FTR. By comparison, standard realignment leaves an ASR of 98.1\%. These results demonstrate that the evaluated adaptive intervention can substantially suppress \tech{} when the defender has access to the exact prefix.

The reduction in attack success is accompanied by lower utility relative to the backdoored model before defense. Under the prefix-aware defense, MMLU-Pro decreases from 47.54 to 42.91, and IFEval prompt-strict decreases from 74.86 to 60.63. When both the prefix and trigger are known, these scores decrease to 42.20 and 53.23, respectively. The IFEval four-metric mean decreases from 81.04 to 67.68 and 61.87, respectively.

Developing and evaluating defenses that explicitly target deferred branching, while preserving model utility and maintaining practical detection costs, is an important direction for future work.